\documentstyle[proceedings,namedreferences]{crckapbk}

\begin{opening}
\title{THE MORPHOLOGICAL EVOLUTION OF FIELD GALAXIES}

\author{RICHARD S ELLIS}
\institute{Institute of Astronomy \\
           Madingley Road, Cambridge CB3 0HA, UK}

\end{opening}

\runningtitle{THE MORPHOLOGICAL EVOLUTION OF FIELD GALAXIES}

\begin{document}

{\small I review two observational programs which,
together, promise to unravel the detailed astrophysical evolution of
normal field galaxies over the last 5-7 Gyr. Systematic ground-based
spectroscopy of faint galaxies have revealed an increasing
faint end slope for the luminosity function with redshift.
The trend is strongest for galaxies undergoing intense
star-formation. Deep images taken with the repaired HST can be
used to count galaxies as a function of morphological type. Regular
`Hubble sequence' galaxies follow the no-evolution prediction,
but irregular/peculiar sources have a steeper count slope and provide
the excess population. Although the overlap between
the spectral and HST samples is currently small, plans to merge
similar datasets should reveal the physical explanation for the
demise of star formation in faint blue galaxies since $z\simeq$0.5-1.}

\setcounter{section}{0}
\section{Introduction}

Multi-object spectroscopy on 4-m telescopes has certainly transformed our
understanding of the faint galaxy population. Controlled surveys to
limiting magnitudes of $b_J$=21.5 (Broadhurst et al 1988), $b_J$=22.5 (Colless
et al 1990,1993) and $B$=24 (Cowie et al 1992, Glazebrook et al 1994) have
revealed a consistent trend in the redshift distribution, $N(z)$, which can
only be logically explained via an increase in the absolute normalisation of
the field galaxy luminosity function (LF) with redshift. The absence of an
excess tail at low $z$ provides a very strong constraint on any uncertainties
in the local LF (c.f. McGaugh 1994). As the blue passband is particularly
sensitive to small changes in star formation, the absence of a dominant high
$z$ tail also restricts strong evolution in massive galaxies. In 1988,
Broadhurst et al proposed that the most likely explanation is a recent decline
in the number and/or star formation rate of sub-L$^{\ast}$ galaxies. They
presented an empirical model which predicts a flattening with time in the
faint end slope of the optical LF.

Babul \& Rees (1992) suggested the faint blue excess is associated with a new
dwarf population, recently `active' in star formation, but subsequently
fading out of view. Broadhurst et al (1990) proposed that distant star forming
`sub-units' may slowly merge to form more massive $L^{\ast}$ galaxies. Both
hypotheses can explain the empirical trends observed, but with different
implications. Only in the latter case is the blue excess clearly related to
the evolutionary history of disk galaxies like the Milky Way.

Might the blue galaxies be morphologically distinguishable from their
`quiescent' counterparts? From the ground, a typical 22nd magnitude galaxy is
a 1.5 arcsecond blur. Despite valiant attempts in sub-arcsecond seeing
(Giraud 1992, Colless et al 1994), the morphologies of typical faint systems
remained unclear. Colless et a) present tentative evidence for the
occurrence of multiple, presumed merging, systems with enhanced star formation
revealed from spectral measures. However, those authors admit the ambiguity of
distinguishing between genuine mergers and irregular systems with patchy H II
regions.

Here I review recent progress on redshift surveys and high resolution imaging
of the faint galaxy population. An extensive new redshift survey, the {\it
Autofib} survey (Heyl 1994, Ellis et al 1994) verifies, for the first time,
that the faint end slope of the LF is steepening with increasing redshift
as predicted by Broadhurst et al (1988). The steepening is attributable
to an increased abundance of star-forming late type galaxies. The repair
of Hubble Space Telescope (HST) allows galaxy counts to be determined {\it
as a function of morphological type}. Again, a population of irregular
systems appears to evolve more rapidly than the remainder. Although the
overlap between the two samples is currently small, it appears likely
that the same subset of the field population is involved.

\section{Uncertainties in the Local Field Galaxy Luminosity Function}

Only recently has the form and absolute normalisation of the local field
galaxy luminosity function (LF) been reliably determined. Efstathiou et al
(1988)
analysed the DARS survey of 326 galaxies in 5 Schmidt-sized fields
to $b_J\simeq$17 demonstrating that a Schechter (1976) form is appropriate.
Without correcting for photometric errors, they determined Schechter
parameters $<M_B^{\ast}, \alpha, \Phi^{\ast}>$ of $<$-19.7, -1.07, 0.0156$>$
($H_o$=100 kms sec$^{-1}$ Mpc$^{-1}$). The more extensive panoramic
sparse-sampled APM-Stromlo southern survey of 1769 galaxies (Loveday et al
1992) found similar parameters $<$-19.7, -1.11, 0.0140$>$. Both surveys
to $b_J$=17 constrain the faint end slope of the local LF only to absolute
magnitudes $M_{bJ}\simeq$-16. An upturn at fainter luminosities such as that
claimed for the Virgo cluster (Binggeli et al 1988) cannot be formally
ruled out. A local population of such feeble sources would have an Euclidean
count slope and might dominate the faint counts diminishing any evolution
that would otherwise be inferred (Kron 1982). Furthermore, surveys limited
at relatively bright apparent magnitudes adopt high surface brightness
detection thresholds and may be poorly-suited for finding intrinsically faint
galaxies (McGaugh 1994). Clearly, the most reliable constraint on the
faint end of the local LF comes from spectroscopy at those faint limits where
the contribution can be directly measured (Glazebrook et al 1994).

A related uncertainty which has plagued the subject concerns the question
of the {\it absolute} normalisation of the LF. Although the DARS and
APM-Stromlo surveys have consistent values of $\Phi^{\ast}$, the number
counts steepen beyond their apparent magnitude limits, 17$<b_J<$20, more than
can be accounted for by non-evolving models, suggesting southern volumes
with $b_J<$17 may be unrepresentative, or dramatic evolution
in recent times (Maddox et al 1990). The explanation for this anomaly
remains unclear but in $\S$5 we present further evidence that local volumes
of radius $\simeq$200 Mpc may be underabundant by a factor of
$\simeq$2.

\section{The Autofib Redshift Survey}

The disparate nature of `benchmark' $b_J<$17 redshift surveys and deep
surveys within narrow faint apparent magnitude slices is not ideal.
Broadhurst et al, Colless et al \& Glazebrook et al were only able to
compare the faint redshift distribution $N(z)$ with empirical
predictions based on the bright survey.  At no redshift was there a
sufficient range in luminosity to examine the form of the LF directly.
Moreover, the number of pencil beams sampled was relatively small ($\simeq$5
each) and some fields are heavily clustered raising the worry that
sampling errors may affect the conclusions.

We have therefore conducted a new `Autofib' survey (using the robotic fibre
positioner built for the Anglo-Australian Telescope -- Parry and Sharples 1988)
spanning a wide apparent magnitude range in many directions and enabling direct
reconstruction of the LF at various redshifts. It includes $\simeq$700
redshifts
from earlier magnitude-limited surveys and about 1000 new redshifts from
Autofib contained within the intermediate magnitude range 17$<b_J<$22
(see Table 1). Further details of the construction and analysis of this new
survey are contained in Heyl's thesis (1994) and preliminary results are
presented here, in Colless (1994) and Ellis et al (1994).

\begin{table}[htb]
\begin{center}
\caption{The Autofib Redshift Survey (Ellis et al 1994)}
\begin{tabular}{lllll}
\hline
Survey & $b_j$ limits & Area deg$^2$& Fields & Redshifts  \\
\hline
DARS(Peterson et al 1986)      & 11.5--16.8 & 70.80 &  5 & 326 \\
BES(Broadhurst et al 1988)     & 20.0--21.5 &  0.50 &  5 & 188 \\
LDSS-1(Colless et al 1990,1993)& 21.0--22.5 &  0.12 &  6 & 100 \\
Autofib bright                 & 17.0--20.0 &  5.50 & 16 & 480 \\
Autofib faint                  & 19.5--22.0 &  4.70 & 32 & 546 \\
LDSS-2(Glazebrook et al 1994)  & 22.5--24.0 &  0.07 &  5 &  73 \\
\hline
TOTAL & & & & 1713 \\
\hline
\end{tabular}
\end{center}
\end{table}

A major difficulty in estimating absolute magnitudes of faint galaxies
given a catalogue of individual $b_J$ magnitudes and redshifts is
estimation of the $k$-correction which depends on the galaxy's (unknown)
spectral energy distribution (SED). At a given redshift within the range
sampled, $k_{bJ}(z)$ changes by $\simeq$1 magnitude across the Hubble
sequence (c.f. King \& Ellis 1985) demonstrating the importance of
inferring the SED of a given faint galaxy.

Broad-band colours are only available for a subset of the survey and
represent a poor substitute for a proper spectral classification
which will be useful in subsequent analyses. Heyl (1994) has devised a
classifier based on cross-correlation of the faint spectra against the wide
aperture local spectral catalogue of Kennicutt (1992). Knowing the
Kennicutt morphology which best matches the faint galaxy, the $k$-correction
is then determined with reference to King \& Ellis' compilation.
Realistic simulations based on Kennicutt's spectra which include photon noise
and sky subtraction difficulties suggest the correct spectral class is
returned to within $\pm$1 class for 90\% of the cases. 6 classes span the
entire Hubble sequence.

One might worry that a class of galaxy exists at faint limits which is not
represented in Kennicutt's list. To check this we devised an {\it internal}
classification scheme based on the [O II] 3727 \AA\ equivalent width and
4000 \AA\ break. Coadding spectra categorised in a 2-dimensional scheme
based on these indices, it is straightforward to identify the
high s/n composites with Kennicutt equivalents, suggesting no serious
omissions. Indeed, with some restrictions, it is even possible to check the
$k$-corrections directly by moving a $b_J$ filter over the coadded data
(Heyl 1994).

Most of the Autofib survey is redshift-complete at the 70-85\% level.
Across the 6 sub-surveys in Table 1, tests show that incompleteness is
primarily a function of apparent magnitude arising from poor continuum
s/n rather than systematics which correlate with galaxy type. This can be
checked by comparing $N(z)$ at the brighter (complete) end of the faint
samples with that for the fainter (incomplete) end of the bright ones.
By weighting in each sample according to the incompleteness at that apparent
magnitude, we can recover satisfactory $V/V_{max}$ distributions for each
spectral class (Colless 1994).

The distribution of absolute magnitudes and redshifts is shown in Figure
1. Luminosity functions have been derived from this data using both traditional
$1/V_{max}$ estimators with errors based on a bootstrap technique and a
step-wise maximum likelihood method modified
as described by Heyl (1994).

\begin{figure}
\vspace{3.0in}  
\caption{Absolute magnitudes and redshifts for the Autofib survey
(Ellis et al 1994). Filled symbols refer to galaxies with strong [O II]
emission.}
\end{figure}

\section{Evolution of the Field Galaxy Luminosity Function}

The enlarged number of faint pencil beams in the Autofib survey leads to new
constraints on the {\it local} LF as well as on its form at high $z$. 560
galaxies in the survey have 0$<z<$0.1 but few are less luminous than
M$_{bJ}\simeq$-16.0; most with $b_J>$22 galaxies lie beyond $z\simeq$0.1.
The paucity of low luminosity galaxies severely limits the size of any
possible upturn in the local LF to $M_{bJ}\simeq$-14 (Figure 2). As the volumes
probed are now quite substantial {\it and} the photometric data used to select
these galaxies penetrate to surface brightness limits below $\mu_{bJ}$=26.5
arcsec$^{-2}$, it becomes hard to argue that the flat LF is due to selection
biases. The redshift distribution at $b_J$=24 eliminates the
possibility that the faint source counts are significantly contaminated by a
population of low luminosity galaxies under-represented in the original
$b_J<$17 surveys (Glazebrook et al 1994).

Figure 2 also shows a highly suggestive steepening of the faint end slope of
the LF with increasing redshift. Formally, a change in shape with $z$ is
significant at the 99.9\% level. Less securely, there is no obvious
brightening in the luminous end of the LF to $z\simeq$1. The latter supports
conclusions derived independently from massive galaxies identified on
the basis of Mg II absorption lines seen in background QSO spectra (Steidel
1994). A picture emerges whereby the LF is composed of two components -- a
luminous one evolving very slowly, if at all, over $z<$1 {\it plus} a
rapidly-evolving component which decays dramatically in the sub-L$^{\ast}$
regime.

\begin{figure}
\vspace{3.0in}  
\caption{Luminosity functions from the Autofib survey (Ellis et al 1994)
in different redshift intervals. Lines refer to Schechter faint end slopes
$\alpha$=-1.1, -1.3 and -1.5.}
\end{figure}

What physical parameters distinguish the galaxies that lie in these two
components? The missing clue appears to be related to the
star-formation rate (c.f. Fig. 1). Figure 3 shows how sources with
strong [O II] emission contribute to the overall absolute magnitude
distribution for the various redshift bins.  The luminosity density of
these star-forming galaxies has decayed by a factor $\simeq$10 since
$z\simeq$0.5.

\begin{figure}
\vspace{4.0in}  
\caption{Absolute magnitude distributions for various redshift
intervals. Shading refers to galaxies whose rest-frame [O II] 3727
\AA\ equivalent width exceeds 20 \AA\ .}
\end{figure}

\section{Faint Galaxy Morphologies from HST}

The repair of HST promises to add a new dimension in the study of faint field
galaxies. Early Cycle 4 images (Griffiths et al 1994) have shown the ease with
which resolved morphological features can be identified in $I\simeq$22
galaxies. A new set of questions can be addressed with such data: (1) What is
the morphological mixture of the high $z$ field population? (2) Are the faint
blue galaxies a distinct morphological population? (3) Can we distinguish
separate evolutionary trends for bulges and disks?

Ideally a large sample of HST morphologies {\it with redshifts} is required
to address these issues. The largest collection of HST images is currently
provided by the `Medium Deep Survey' (PI: Griffiths). In this key project,
WFPC-2 is used in parallel mode associated with primary pointings defined by
a variety of other observers. Redshifts have to be secured later. A limited
amount of HST imaging has been done in the reverse mode: i.e. primary WFPC-2
imaging of fields with existing spectroscopy from the {\it Autofib} survey
(PI: Broadhurst). Figure 4 shows how powerful the combination of
ground-based spectroscopy and HST imaging will be. In a single WFPC-2 image
10 $b_J<$24 galaxies from the LDSS-1/2 surveys reveal morphological
types in excellent agreement with their spectral classes.
\begin{figure}
\vspace{65mm}  
\caption{Ground-based spectrum and WFPC-2 image of a faint
galaxy from the HST program of Broadhurst et al. The late-type
spectral class is confirmed morphologically.}
\end{figure}

The most significant result to date comes from the
Medium Deep Survey. Over 300 $I<$22 galaxies have been classified
on a simple E/S0: Spiral: Irr/Pec scheme (Glazebrook et al 1995).
The number-magnitude counts as a function of type are shown in Figure 5
and illustrate that the spiral and early-type classes show little
evidence for evolution. Significantly, both classes fit the predictions
only if the absolute normalisation is $\times\simeq$2 higher than that
derived from the 17th magnitude surveys (c.f. $\S$2). On the other hand,
the irregular and peculiar galaxies demonstrate a count slope much
steeper than expected, consistent with significant evolution. Although the
overlap with the spectral samples remains small, it seems highly likely
that the [O II]-strong sources which decline dramatically in number
since $z\simeq$1 are the morphologically unusual examples in the HST
samples. A morphologically-distinct population of sources appears to be
responsible for the well-established excess population of faint blue galaxies.

\begin{figure}
\vspace{3.5in}  
\caption{Morphological counts from the HST Medium Deep Survey
(Glazebrook et al 1995). Those of `regular' Hubble types are consistent
with no evolution but irregular/peculiars show a steep slope suggesting
rapid evolution.}
\end{figure}

\noindent{\bf Acknowledgements:} The {\it Autofib} survey involves
Matthew Colless, Tom Broadhurst, Jeremy Heyl and Karl Glazebrook and
the Medium Deep Survey is led by Richard Griffiths. I thank all
co-workers for allowing me to present data prior publication. I
acknowledge financial support from the IAU and the assistance of
Jacqueline Bergeron and Piet van der Kruit.


\begin{thebibliography}{}
\bibitem{}
Babul, A. \& Rees, M.J. (1992), {\it MNRAS}, {\bf 255}, 346.
\bibitem{}
Binggeli, B., Sandage, A. \& Tammann, G. (1988), {\it Ann. Rev. Astron.
Astr.},{\bf 26}, 509.
\bibitem{}
Broadhurst, T.J., Ellis, R.S. \& Shanks, T. (1988),
{\it MNRAS}, {\bf 235}, 827.
\bibitem{}
Broadhurst, T.J., Ellis, R.S. \& Glazebrook, K. (1992),
{\it Nature}, {\bf 355}, 55.
\bibitem{}
Colless, M., Ellis, R.S., Taylor, K. \& Hook, R. (1990),
{\it MNRAS}, {\bf 244}, 408.
\bititem{}
Colless, M., Ellis, R.S., Broadhurst, T.J., Taylor, K. \&
Peterson, B.A. {\it MNRAS}, {\bf 261}, 19 (1993).
\bibitem{}
Colless, M., Schade, D., Broadhurst, T.J. \& Ellis, R.S. (1994),
{\it MNRAS}, {\bf 267}, 1108.
\bibitem{}
Colless, M. in {\it Wide Field Spectroscopy}, eds. Maddox, S.J.
\& Arag\'on-Salamanca, A., World Scientific, in press.
\bibitem{}
Cowie, L., Songaila, A. \& Hu, E.M. (1991), {\it Nature}, {\bf 354}, 460.
\bibitem{}
Efstathiou, G., Ellis, R.S. \& Peterson, B.A. (1988),
{\it MNRAS}, {\bf 232}, 431.
\bibitem{}
Ellis, R.S., Broadhurst, T.J, Colless, M.M, Heyl, J.S. \& Glazebrook,
K. (1994), {\it MNRAS}, submitted.
\bibitem{}
Giraud, E. (1992), {\it Astron. Astrophys.}, {\bf 257}, 501.
\bibitem{}
Glazebrook, K., Ellis, R.S., Colless, M., Broadhurst, T.J.,
Allington-Smith, J. \& Tanvir, N. (1994), {\it MNRAS}, in press.
\bibitem{}
Glazebrook, K., Ellis, R.S., Santiago, B. \& Griffiths, R. (1995),
{\it Nature}, submitted.
\bibitem{}
Griffiths, R.E. {\it et al} (1994) {\it Ap. J. Lett.}, {\bf 435}, L19.
\bibitem{}
Heyl, J. 1994 M.Sc. thesis, University of Cambridge, UK.
\bibitem{}
Kennicutt, R. C. (1992), {\it Ap.J.}, {\bf 388}, 310.
\bibitem{}
King, C.R. \& Ellis, R.S. (1985), {\it Ap.J.}, {\bf 288}, 456.
\bibitem{}
Kron, R. (1982), {\it Vistas}, {\bf 26}, 37.
\bibitem{}
Loveday, J., Peterson, B.A., Efstathiou, G., Maddox, S.J. \&
Sutherland, W.J. (1992), {\it Ap. J.}, {\bf 390}, 338.
\bibitem{}
Maddox, S.J., Sutherland, W.J., Efstathiou, G., Loveday, J.
\& Peterson, B.A. (1990), {\it MNRAS}, {\bf 247}, 1P.
\bibitem{}
McGaugh, S. (1994) {\it Nature}, {\bf 367}, 538.
\bibitem{}
Parry, I.R. \& Sharples, R.M. (1988), in {\it Fiber Optics in
Astronomy}, ed. Barden, S.M., PASP series Vol. 3, p93.
\bibitem{}
Steidel, C. (1994) in {\it Wide Field Spectroscopy}, eds. Maddox, S.J.
\& Arag\'on-Salamanca, A., World Scientific, in press.
\end{thebibliography}
\end{document}